\documentclass[prl,showpacs,preprint]{revtex4-1} 
\usepackage{xspace} 
\usepackage{makecell} %\thead
\usepackage{siunitx, mhchem}
\usepackage{color,soul}
\usepackage{graphicx}
\usepackage{bm}        % for math
\usepackage{amssymb}   % for math\usepackage{amsfonts}
\usepackage{amsmath,mathtools} % mathtools to use \begin{multlined}
\usepackage{breqn} %Break Equation automatically
\usepackage{multirow}
\usepackage{array}
\usepackage{booktabs}
\usepackage[export]{adjustbox}
\usepackage{textcomp}%for \textdegree
\usepackage[normalem]{ulem} %To use \sout strikout
\usepackage{gensymb}
\newcommand{\mos}{MoS$_2$\xspace}

\begin{document}
%%%%%%%%%%%%%%%%%%%%%%%%%%%%%%%%%%%%%%%%%%%%%%%%%%%%%%%%%%%%
%\newcommand{\taum}{\tau_\text{\tiny M}}
%\documentclass[journal]{paper}
%\documentclass[aip,apl,reprint]{revtex4-1}
%%%%%%%%%%%%%%%%%%%%%%%%%%%%%%%%%%%%%%%%%%%%%%%%%%%%%%%%%%%
%\usepackage{xspace} 
%\usepackage{cite} %1-3,5-7 instead 1,2,3,5,6,7
%\usepackage{makecell} %\thead
%\usepackage{siunitx, mhchem}
%\usepackage{color,soul}
%\usepackage{graphics}
%\usepackage{bm}        % for math
%\usepackage{amssymb}   % for math
%\usepackage{amsfonts}
%\usepackage{amsmath,mathtools} % mathtools to use \begin{multlined}
%\usepackage{breqn} %Break Equation automatically
%\usepackage{multirow}
%\usepackage{array}
%\usepackage{booktabs}
%\usepackage[export]{adjustbox}
%\usepackage{textcomp}%for \textdegree
%\usepackage[numbers]{natbib}
%\usepackage{epstopdf}
%\usepackage{gensymb}
%\DeclareGraphicsExtensions{.eps}
%%%%%%%%%%%%%%%%%%%%%%%%%%%%%%%%%%%%%%%%%%%%%%%%%%%%%%%%%%%%
%\newcommand{\mos}{MoS$_2$\xspace}
%\newcommand{\ws}{WS$_2$ }
%\newcommand{\degree}{$^{\small{\text{o}}}$}
%%%%%%%%%%%%%%%%%%%%%%%%%%%%%%%%%%%%%%%%%%%%%%%%%%%%%%%%%%%%

%\begin{document}
	
	% Use the \preprint command to place your local institutional report
	% number in the upper righthand corner of the title page in preprint mode.
	% Multiple \preprint commands are allowed.
	% Use the 'preprintnumbers' class option to override journal defaults
	% to display numbers if necessary
	%\preprint{}
	
	%Title of paper
	\title{Effect of manganese incorporation on the excitonic recombination dynamics in monolayer \mos}

	\author{Poulab\ Chakrabarti}
	\email{poulab007@gmail.com}
	\author{Santosh\ Kumar\ Yadav}
	\author{Swarup\ Deb}
	\author{Subhabrata\ Dhar}
	%\email{dhar@phy.iitb.ac.in}
	\affiliation{Department of Physics$,$ Indian Institute of Technology Bombay$,$ Powai$,$ Mumbai 400076$,$ India}

	\begin{abstract}
		 Using X-ray photoelectron spectroscopy (XPS), atomic force microscopy (AFM) and Raman spectroscopy techniques we investigate the incorporation of Manganese (Mn) in monolayer (1L)-\mos grown on sapphire substrates by microcavity based chemical vapor deposition (CVD) method. These layers are coated with different amount of Mn by pulsed laser deposition (PLD) technique and temperature dependent photo-luminescence (PL) spectroscopic study has helped us in understanding how such deposition affects the dynamics of excitonic recombination in this system.  The study further  reveals two distinctly different  Mn-incorporation regimes. Below a certain critical deposition amount of Mn, thin Mn-coating with large area coverage is found on \mos layers and in this regime, substitution of Mo ions by Mn is detected through XPS. Dewetting takes place when Mn-deposition crosses the critical mark, which results in the formation of Mn-droplets on \mos layers. In this regime, substitutional incorporation of Mn is suppressed, while the Raman study suggests an enhancement of disorder in the lattice with the Mn-deposition time. From PL investigation, it has been found that the increase of the amount of Mn-deposition not only enhances the density of non-radiative  recombination channels for the excitons but also raises the barrier height for such recombination to take place. The study attributes these non-radiative transitions to certain Mo related defects (either Mo-vacancies or distorted Mo-S bonds), which are believed to be generated in large numbers during Mn-droplet formation stage as a result of the withdrawal of Mn ions from the Mo-substitutional sites.
	\end{abstract}
	
	% insert suggested PACS numbers in braces on next line
%	\pacs{}
	% insert suggested keywords - APS authors don't need to do this
	%\keywords{}
	
	%\maketitle must follow title, authors, abstract, \pacs, and \keywords
	\maketitle
	
	% body of paper here - Use proper section commands
	% References should be done using the \cite, \ref, and \label commands

%\section{Introduction}
%\vspace*{-\baselineskip}
Monolayer (1L) \mos, a two dimensional (2D) transition metal dichalcogenide (TMD), has become one of the most studied materials in the last decade for its unique electronic and optical properties\cite{mak1}, among which spin-valley coupling\cite{mak2,VP,VP1} and direct band gap at the K-points of the Brillouin zone\cite{mak1,bs} are the most important ones. Because of its direct band gap of $\sim$1.9 eV and large exciton binding energy, 1L-\mos offers high luminescence yield in the visible range\cite{mak1,bs,VP1}. Atomically thin size and flexible nature make the material a potential candidate for 3D integrated and flexible electronics of the future \cite{3D_flex}. Furthermore, there are multiple theoretical proposals that predict ferromagnetic behaviour even above room temperature in 1L-\mos, when Mo is substituted by certain transition metal ions. Such a two dimensional dilute magnetic semiconductor (2D-DMS) has potential for injecting spin polarized excitons and carriers for future 2D spintronic and valleytronic devices\cite{magnet1,magnet2,magnet3,VP2}. Achieving controllable doping is one of the important requirements for any technology application of the material. However, doping of 1L-\mos is always onerous because of the strong bonding between Mo and S, which makes it almost inert for any substitutional reaction to take place\cite{XPS1,bs}. This has become even more challenging as the unambiguous detection of doping is extremely difficult in an atomically thin layer. Even though, there are efforts by various groups to incorporate different elements in 1L-\mos using varied  means, such as chemical doping\cite{ch1,ch2,ch3},  physisorption\cite{therm,ad2,ad3,ad4}, metal evaporation\cite{evap1,raman2} and ion irradiation\cite{ion1,raman1}, a method for stable and sustainable doping of the layer is yet to be established.  Previously, Zhang \textit{et al.} have reported up to 2 atomic percentage of  Manganese (Mn) doping in 1L-\mos by vapor phase deposition technique\cite{XPS2}. Mn incorporation has also been tried using Mn ion irradiation technique\cite{raman1}, which has been found to result in a large amount of structural defects in the lattice. More studies are needed to understand the mechanism of incorporation of Mn and other metals in 1L-\mos. It is also important  to investigate how the incorporation can influence the electrical, optical and magnetic properties of the layer.             

Here, we present a detailed study of manganese doping in 1L-\mos through pulsed laser deposition (PLD) technique and its influence on the optical properties of the films, which are grown on sapphire substrates by microcavity based chemical vapor deposition (CVD) technique\cite{pkm,tailoring}. Our investigation using high resolution X-ray photoelectron spectroscopy (XPS), atomic force microscopy (AFM), Raman and PL spectroscopy reveals two clearly distinct regimes of Mn-incorporation in 1L-\mos. When the deposition amount is below a certain critical value, Mn islands covering large areas are found to be formed on \mos layers and substitution of Mo ions by Mn are observed. As the deposition amount goes beyond the critical value, dewetting leads to the formation of Mn-droplets on \mos layers. In this regime, substitutional incorporation of Mn is decreased, while the disorder in the lattice increases with the deposition time.  PL study shows that Mn-deposition strongly influence the excitonic recombination dynamics by opening up non-radiative recombination channels. Both the density and the barrier height for such recombinations are found to increase  with the amount of Mn-deposited.  These non-radiative transitions are attributed to certain Mo related defects. 

%\section{Experimental Techniques}

Strictly monolayer (1L) \mos films were grown on \textit{c}-sapphire substrates by the microcavity based CVD route where molybdenum trioxide (MoO$_3$) and sulfur (S) powders were used as precursors and argon (Ar) was used as the carrier gas.  More details about the growth can be found elsewhere\cite{pkm,tailoring}. Manganese (Mn) was deposited on these layers using pulsed laser deposition (PLD) technique where a polycrystalline Mn pellet was used as target. The pellet was prepared from 99.99\% pure Mn powder by arc melting technique in Ar (5N pure) atmosphere. Substrate (1L-\mos/sapphire) temperature was maintained at 300$\degree$C during the deposition. Distance between the Mn-target and the sample was kept at 5 cm for all depositions. A KrF excimer laser of wavelength 248 nm and pulse width of 25 ns was used for ablating the target. Deposition amount was varied by varying the laser pulse count. While the pulse frequency and the energy density of the laser were maintained at 1 Hz and 1.4 J/cm$^2$, respectively, for all cases. Base pressure inside the PLD chamber was kept at $\sim$ 8.0$\times$10$^{-5}$ mbar. During the growth, 5N argon gas was introduced inside the chamber to maintain a pressure of  $\sim$ 8.0$\times$10$^{-2}$ mbar inside the chamber. 1L-\mos layers were coated with Mn for different numbers of laser pulse count ranging from 1 to 10. These samples are labeled by alphabets A to D. In  another sample (sample E), Mn deposition was carried out at the base pressure  ($\sim$ 8.0$\times$10$^{-5}$ mbar) for 10 counts of laser pulses.  Mn was also deposited on a bare \textit{c}-plane sapphire substrate (without \mos) for five counts of  laser pulses at 8.0$\times$10$^{-2}$ mbar of Ar-pressure, keeping other parameters constant. This sample is named as sample F. One as-grown 1L-\mos sample R, was used as the reference standard for all the measurements. Mn deposition parameters for these samples, are listed in Tab.~\ref{tab1}. 

X-ray photoelectron spectroscopy on these samples were performed in Axis Supra spectrometer from Kratos Analytical, UK, where monochromatic Al K${_\alpha}$ x-ray (h$\nu$ = 1486.6~eV) was used as excitation source. Energy axis of the recorded spectra, was calibrated with respect to the binding energy of carbon 1$s$ peak at 284.7~eV. Morphology of the samples were examined by atomic force microscopy (AFM) in tapping mode using  Bruker, NanoScope-IV system. Raman spectra on these monolayers were recorded at room temperature in back scattering geometry with 532~nm diode laser using Horiba Jobin Yvon HR800 confocal Raman spectrometer. A closed-cycle helium refrigerator was used to carry out temperature dependent (from 20~K to 300~K) photoluminescence (PL) measurements, where 532~nm diode laser was used as the excitation source and the spectra were recorded by a 0.55~m focal length monochromator attached with a Peltier cooled CCD camera.

\begin{center}
	\begin{table}
		\caption{Mn deposition parameters for different samples. Substrate temperature, distance between target and substrate, frequency and energy density of the laser are maintained at 300$\degree$C, 5~cm, 1~HZ and 1.4 J/cm$^2$, respectively for each deposition.}
		\begin{tabular}{c| c |c |c}
			
			\hline
			{\small Sample} & {\small Pressure} & {\small Ar flow rate} & {\small Laser Pulse count} \\ 
			& {\small(mbar)} & {\small(sccm)} &  \\
			\hline
			\hline	
			~R~& -- & -- & 0  \\    
			%			\hline
			~A~& {\small 8.0$\times$10$^{-2}$} & 40 & 1  \\    
			%			\hline
			~B~& {\small 8.0$\times$10$^{-2}$} & 40 & 5  \\
			%			\hline 
			~C~& {\small 8.0$\times$10$^{-2}$} & 40 & 7  \\
			%			\hline
			~D~& {\small 8.0$\times$10$^{-2}$} & 40 & 10  \\
			%			\hline
			~E~& {\small 8.0$\times$10$^{-5}$} & 0 & 10  \\
			%			\hline
			~F (Sapphire)~& {\small 8.0$\times$10$^{-2}$} & 40 & 5  \\
			\hline            		    
		\end{tabular}
		\label{tab1}
	\end{table}
\end{center}

%\section{Results and Discussion}

\begin{figure}
	\includegraphics[scale=1.8]{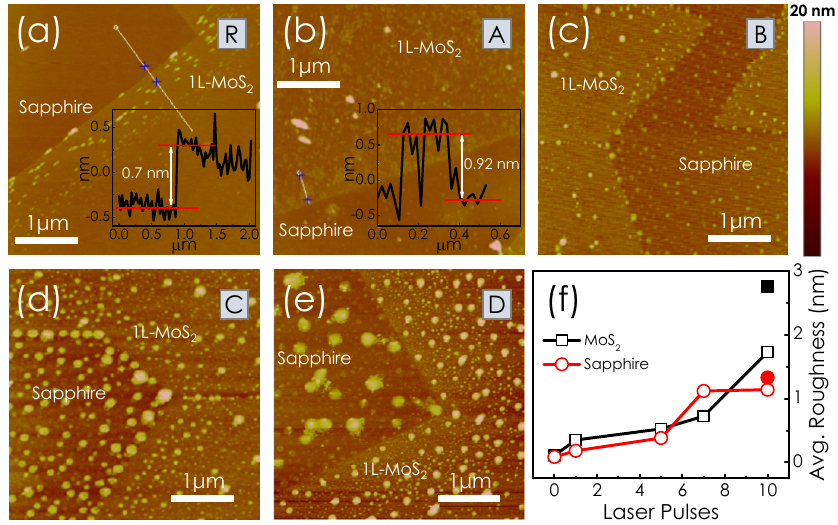}
	\caption{AFM images for samples (a) R, (b) A, (c) B, (d) C and (e) D. Inset of (a) shows the height distribution along a line drawn across the boundary. Inset of (b) shows the line scan profile over a Mn-deposited island on \mos film. (f) Average roughness of \mos (black) and sapphire (red) regions as functions of pulse count. Filled symbols represent sample E.}
	\label{fig1}
%	\vspace*{-\baselineskip}
\end{figure}

Fig.~\ref{fig1} shows AFM images for the \mos samples coated with Mn for different counts of laser pulses. In all the images, \mos layers can clearly be distinguished from the bare sapphire surface from the height contrast. As shown in Fig.~\ref{fig1}(a), average step-height for the reference sample R has been found to be $\sim$ 0.7~nm, which is expected for a monolayer \mos\cite{pkm,tailoring}. In fact, it should be noted that average step-height for all samples before Mn deposition is found to be $\sim$ 0.7~nm confirming monolayer nature of \mos in all cases. As shown in Fig.~\ref{fig1}(b), deposition of several flat island type structures with average height of $\sim$ 0.9~nm on the \mos regions is evident in sample A, where only one laser pulse is used for Mn deposition. In case of sample B, where Mn is deposited for 5 laser pulses, flat island type structures could not be distinguished clearly on \mos regions, instead, several spherical structures are visible on both \mos and bare sapphire regions as shown in Fig.~\ref{fig1}(c). Density and size of the spherical droplets clearly increase as one moves from sample C [Fig.~\ref{fig1}(d)]  to D [Fig.~\ref{fig1}(e)], meaning as the laser pulse count for Mn deposition increases (see supplementary Fig.~S1\cite{sup}). Noticeably, in these samples, droplets on bare sapphire are bigger than those on \mos regions. It is noteworthy that similar type of droplet formation has also been observed in case of sample E (see supplementary Fig.~S1\cite{sup}). Fig.~\ref{fig1}(f) plots the average value of the roughness of \mos (black symbols) and sapphire (red symbols) regions as functions of laser pulse count. Evidently, variation of roughness on \mos regions shows a clear increase beyond 5 counts of laser pulse. Note that the average roughness on sapphire regions also shows a sudden jump beyond 5 pulses. These are consistent with the sudden change in the morphology observed in the AFM images and seem to suggest that Mn-deposition for sufficiently less number of laser pulses results in a thin Mn-layer. The strain/surface energy developed in such a thin metal film may not be sufficient to change its morphology when the sample is cooled to room temperature after deposition. It is plausible that in case of the samples, where Mn-deposition is performed for the number of pulses more than a certain value, thickness of the deposited Mn-film crosses a critical mark and the strain/surface energy developed in the layer becomes so high that to minimise energy, the film breaks into Mn-droplets when the sample is brought to the room temperature. This point has been discussed again later (in Fig.~\ref{fig5}).

\begin{figure*}[t]
	\includegraphics[scale=1]{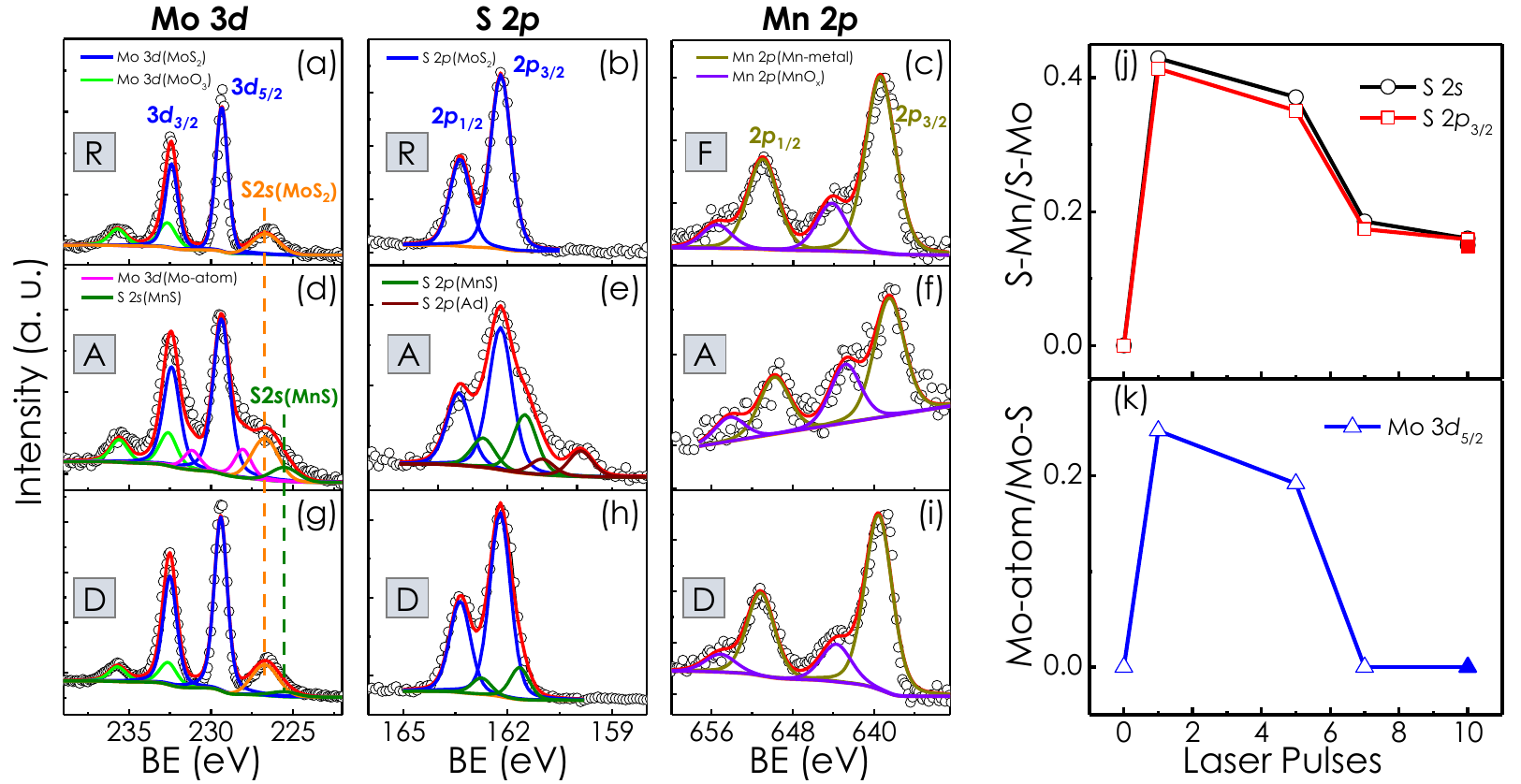}
	\caption{XPS core level spectra for Mo $3d$ and S $2s$ levels  in case of samples (a) R, (d) A and (g) D; for S $2p$ levels  in case of the samples (b) R, (e) A and (h) D; for Mn $2p$ levels in case of  the samples (c) F,  (f) A and (i) D. (j) Ratio of the contributions from S-Mn and S-Mo bonds obtained by de-convoluting S~$2s$ (black symbols) and S~$2p_{3/2}$ (red symbols) spectra for these samples versus pulse count. (k) Ratio of the contribution from Mo atoms and Mo-S bonds obtained from the Mo~$3d_{5/2}$ profiles versus pulse count. In (j) and (k), filled symbols represent sample E.}
	\label{fig2}
%	\vspace*{-\baselineskip}
\end{figure*}

In panels (a-i) of Fig.~\ref{fig2}, XPS spectra for Mo~$3d$, S~$2s$, S~$2p$ and Mn~$2p$ levels for different samples are compared. These spectra are deconvoluted with Voigt functions as mentioned in earlier references\cite{XPS,XPS1}. In all panels, red lines represent the fitted profiles. For the reference sample R [panel (a)], S~$2s$ (orange line) peak  appears (at 226.7~eV) along with Mo~$3d$ doublets. Mo~$3d$ related features can be deconvoluted into two doublets. While, the position of the first doublet, which peaks at 229.3~eV (Mo $3d_{5/2}$) and 232.4~eV (Mo $3d_{3/2}$) [blue line], matches quite well with Mo-S bonds in \mos\cite{XPS,XPS1}, the second doublet appearing at  232.6~eV (Mo~$3d_{5/2}$), 235.7~eV (Mo~$3d_{3/2}$) [bright green line] can be assigned to the Mo-O bonds of the unreacted MoO$_3$ particles on the surface\cite{XPS}. Panel (b) shows the S $2p$ features for the reference sample. Deconvolution of the feature results in two peaks centred at  162.2~eV and  163.4~eV, (blue line) which are associated with S~$2p_{3/2}$  and S~$2p_{1/2}$  lines, respectively,  of \mos\cite{XPS,XPS1}. Fig.~\ref{fig2}(c) shows Mn $2p$ features for sample F, which is a bare sapphire substrate coated with manganese. The feature can be deconvoluted into two doublets. While the doublet (olive line) appearing at  639.2~eV (Mn~$2p_{3/2}$) and  651.0~eV  (Mn~$2p_{1/2}$) can be attributed to Mn-metal\cite{XPS2,XPS3}, position of the higher energy doublet (purple line) appearing at 644.0~eV (Mn~$2p_{3/2}$) and 655.6~eV (Mn~$2p_{1/2}$) matches quite well with those of Mn-O bonds of manganese oxide (MnO$_x$)\cite{XPS2,XPS5}. Quite evidently, Mo~$3d$, S~$2s$ as well as S~$2p$ features appears much broader for sample A as compared to the reference sample as can be seen in  Fig.~\ref{fig2}(d) and (e). This is a clear indication of the change of the bonding states of both Mo and S ions in this sample. Deconvolution of these features results in the formation of one additional  Mo $3d$ doublet peaking at  227.9~eV (Mo~$3d_{5/2}$)  and  231.0~eV (Mo~$3d_{3/2}$) [magenta line in Fig.~\ref{fig2}(d)], one extra S $2s$ peak (at 225.5~eV) [bottle green line in Fig.~\ref{fig2}(d)] and two extra S $2p$  doublets; one peaking at 161.4~eV (S~$2p_{3/2}$)  and 162.6~eV (S~$2p_{1/2}$) [bottle green line in Fig.~\ref{fig2} (e)] and the other, peaking at 159.9~eV (S~$2p_{3/2}$)  and 161.0~eV (S~$2p_{1/2}$)[brown line in Fig.~\ref{fig2}(e)]. Position of the new Mo~$3d$ doublet [magenta line in Fig.~\ref{fig2} (d)] matches well with neutral Mo atoms\cite{XPS1}. Out of the two extra S~$2p$ doublets, the bottle green line of Fig.~\ref{fig2}(e) can be assigned to S-Mn bond formation and the brown line of Fig.~\ref{fig2}(e) can be attributed to surface adsorbed S atoms\cite{XPS1}. The extra S~$2s$ peak at 225.5~eV [bottle green line in Fig.~\ref{fig2}(d)] can also be attributed S-Mn bonds\cite{XPS4,XPS2}. Interestingly for sample D, Mo~$3d$, S~$2s$ and S~$2p$ profiles are similar to those obtained for the reference sample as shown in Fig.~\ref{fig2}(g) and (h). Deconvolution of these profiles results in a much weaker contributions from the S-Mn bonds and neutral Mo atoms as compared to what has been found in case of sample A. These findings point towards the fact that for sample A, Mn atoms partially substitute Mo ions of \mos to form Mn-S bonds and the knocked out Mo ions settle as neutral Mo atoms. While, in sample D, such substitution is very weakly present. It should be noted that like in sample A, significant presence of additional Mo and S related features could also be found in sample B, where Mn is deposited for 5 laser pulses. Similar to the case of sample D, very weak sign of Mo-substitution by Mn-ions could be found for sample C and E, where laser pulse counts of 7 and 10, respectively, are used for Mn deposition. Note that Mo~$3d$, S~$2s$ and S~$2p$ spectra for all the Mn deposited \mos samples are shown in supplementary figure S2\cite{sup}. Fig.~\ref{fig2}(f) and (i) show the  Mn $2p$ features for sample A and D, respectively. It is interesting to note that for both the cases, features are very similar to what has been found for sample F [Fig.~\ref{fig2}(c)] showing the contributions only from Mn-metal and Mn-O bonds. In fact, no sign of Mn-S bond formation could be found from Mn-spectra for any of the Mn-coated \mos samples. Note that in all these samples, a significant portion of the sapphire surface remains uncovered by 1L-\mos. Perhaps, the signal due to Mn-S bond from the Mn-coated \mos layers is overwhelmed by the signal arising from the Mn-coated sapphire regions for all these samples. The ratio of the contributions from S-Mn and S-Mo bonds can be obtained in two ways; by deconvoluting (i) S~$2s$ and (ii) S~$2p_{3/2}$ spectra. These results are plotted as a function of the laser pulse count in Fig.~\ref{fig2}(j). Evidently, both the profiles show very similar trend. Similarly, the ratio of the contribution from Mo atoms and Mo-S bonds can be achieved from the Mo~$3d_{5/2}$ peaks, which is plotted versus pulse count in Fig.~\ref{fig2}(k). Interestingly, no signature of knocked out Mo atoms could be found in samples deposited with more than 5 pulse counts. It is quite evident from these plots that substantial amount of Mo-substitutional incorporation of Mn in 1L-\mos layers is achieved only for pulse counts of 1 and 5. Incorporation sharply decreases as the pulse count  goes beyond 5.                       

\begin{figure}
	\includegraphics[scale=1.8]{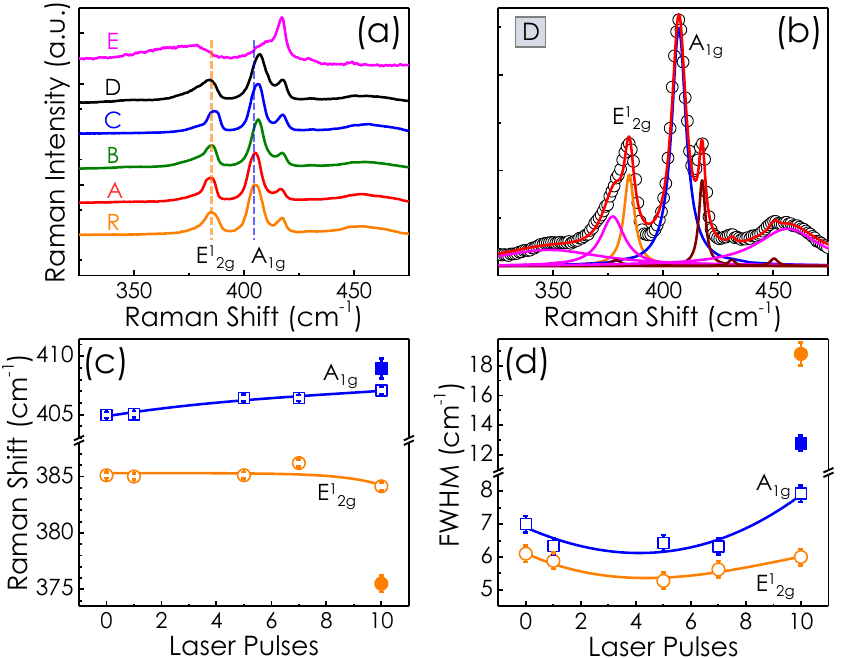}
	\caption{(a) Room temperature Raman spectra for different Mn-deposited samples in the frequency range containing characteristic first order E$^1_{2g}$ and A$_{1g}$ peaks of \mos. (b) Deconvolution of the spectrum with Lorentzian functions for sample D. (c) Peak position and (d) FWHM  of the Lorentzians associated with E$^1_{2g}$ (orange) and A$_{1g}$ (blue) peaks versus pulse count. Filled symbols represent sample E. Solid lines in (C) and (d) are guides to the eyes.}
	\label{fig3}
%	\vspace*{-\baselineskip}
\end{figure}

Fig.~\ref{fig3}(a) compares room temperature micro-Raman spectra for these samples in the frequency range, where characteristic in-plane E$^1_{2g}$ and out-of-plane A$_{1g}$ vibrational modes for the zone-centre phonons are observed at $\sim$ 385 and 405~cm$^{-1}$, respectively. These spectra are deconvoluted with Lorentzian functions as shown in  the Fig.~\ref{fig3}(b) for sample D as an example. Apart from the  dominant E$^1_{2g}$  (orange line) and A$_{1g}$ (blue line) features, several other phonon modes associated with M and K points of the Brillouin zone (BZ) [magenta lines] can also be seen. Note that these non-zone-centre phonons become visible due to the relaxation of the fundamental Raman selection rule (phonon wave vector $q$ $\approx$ 0) in the presence of disorder in the MoS$_2$ lattice (see supplementary Fig.~S3\cite{sup})\cite{raman1,raman2, raman4,raman5}. There are also a couple of other features (brown lines) in the spectrum, which can be identified as the Raman lines arising from the sapphire substrate (see the supplementary Fig.~S3\cite{sup}). Fig.~\ref{fig3}(c) and (d) plot the peak position and full width at half maximum (FWHM) of the Lorentzians associated with E$^1_{2g}$ (orange) and A$_{1g}$ (blue) peaks as functions of the number of laser pulses used for Mn-deposition in these samples. It is evident from these figures that with the increase of pulse count,  A$_{1g}$  peak shows a tendency of shift towards higher wavenumbers.  On the other hand,  position of E$^1_{2g}$ peak does not vary much with pulse count upto 7, beyond which it reduces. It should be noted that electron concentration in the conduction band has a strong influence on the coupling between electrons and A$_{1g}$ phonons in 1L-\mos\cite{raman3}. Increase of electron concentration in the layer is found to result in the softening of frequency as well as broadening of the A$_{1g}$ peak, while E$^1_{2g}$ feature hardly shows any change\cite{raman1,raman2,raman3}. Note that the FWHM of the A$_{1g}$ peak is less than that of the reference sample for samples A and B, where the pulse counts are 1 and 5, respectively. This may suggest a reduction of electron concentration in the two \mos samples, which can further imply that Mn-ions incorporate in \mos when the number of laser pulses used for Mn deposition is sufficiently less.  This observation is also consistent with the results of the XPS study shown in Fig.~\ref{fig2}. Increasing trend of the line-width for both A$_{1g}$ and E$^1_{2g}$ peaks as well as hardening and softening of A$_{1g}$ and E$^1_{2g}$ modes, respectively, for pulse count beyond 7 may indicate a rise of disorder in the layer with the increase of Mn-deposition time.  

\begin{figure}
	\includegraphics[scale=1.8]{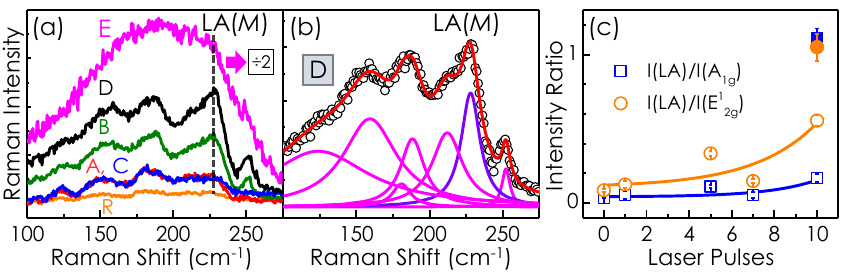}
	\caption{ (a) Lower energy portion of the room temperature Raman spectra for various Mn-deposited samples. (b) Deconvolution of the structure with Lorentzian functions for sample D. Note that the most dominant peak at $\sim$ 227~cm$^{-1}$  has been assigned to longitudinal acoustic phonon mode at \textit{M} point [LA(\textit{M})]. (c) Intensity ratio between LA(\textit{M}) and A$_{1g}$ peaks [I(LA)/I(A$_{1g}$)] as well as LA(\textit{M}) and E$^1_{2g}$ peaks [I(LA)/I(E$^1_{2g}$)] versus pulse count. Filled symbols represent sample E. Solid lines provide the visual guidance.}
	\label{fig4}
%	\vspace*{-\baselineskip}
\end{figure}

Fig.~\ref{fig4}(a) highlights the lower energy portion of the Raman spectra for these samples, where certain zone edge phonons give rise to broad structure with multiple features due to the presence of disorder\cite{raman1,raman2,raman4,raman5}. Fig.~\ref{fig4}(b)  shows the Lorentzian deconvolution of the structure for sample D, where the most dominant peak at $\sim$ 227~cm$^{-1}$ (violet line) has previously been assigned to longitudinal acoustic phonon mode at \textit{M} point [LA(\textit{M})]. This peak has also been treated as a measure of disorder in 1L-\mos films\cite{raman1} . Please see supplementary Fig.~S3\cite{sup} for identification of other features in the structure. It should be noted that the spectra shown in all the above panels are normalized with respect to the intensity of A$_{1g}$ peak.  From Fig.~\ref{fig4}(a) it is evident that all the disorder related peaks  enhance with the laser pulse count suggesting a gradual increase of disorder with the amount of Mn deposition in these samples. In case of sample E, where Mn-deposition is carried out for 10 pulse-counts in vacuum (switching off the Ar supply), peak-shift and peak-width for both E$^1_{2g}$ and A$_{1g}$ peaks are dramatically higher than the other samples suggesting much higher degree of disorder in this sample [see Fig.~\ref{fig3}(c) and (d)].  Intensity ratios between LA(\textit{M}) and A$_{1g}$ peaks [I(LA)/I(A$_{1g}$)] (blue line) as well as  LA(\textit{M}) and E$^1_{2g}$ peaks [I(LA)/I(E$^1_{2g}$)] (orange line), are plotted in Fig.~\ref{fig4}(f). A gradual rise of disorder with increasing Mn-deposition time is quite evident from the figure. This observation is in line with the results shown in Fig.~\ref{fig3}.

\begin{figure}
	\includegraphics[scale=1.8]{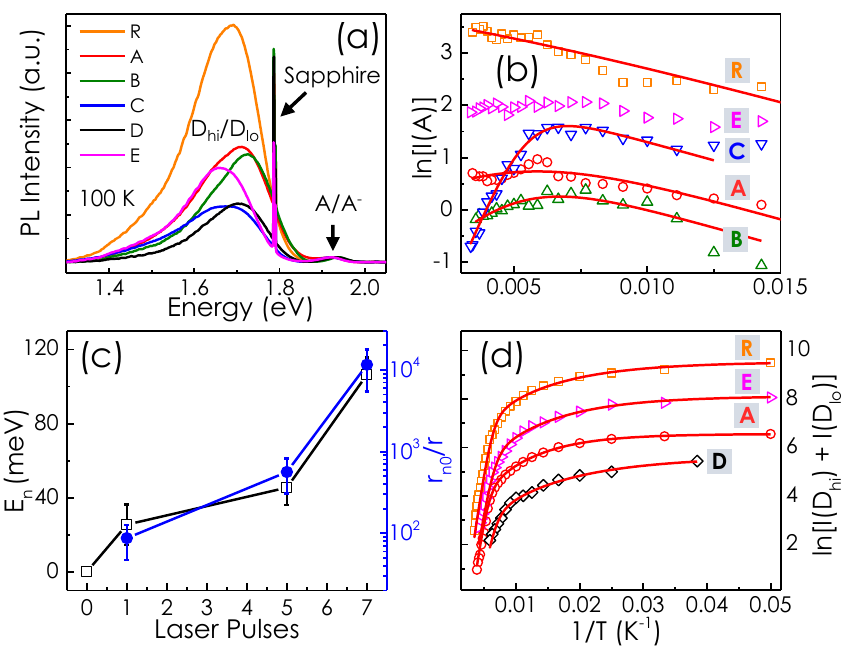}
	\caption{(a) Photoluminescence (PL) spectra  recorded at 100~K for various Mn-deposited samples. (b) Log of the integrated intensity associated with A-exciton [$I(A)$] versus inverse of temperature. (c) Thermal energy barrier $E_n$ and $r_{n0}/r$ for non-radiative recombination, which are obtained from the fitting as described in the text, versus pulse count. (d) Logarithm of the intensity of the BL feature as a function of the inverse of temperature for different Mn-deposited samples.}
	\label{fig5}
%	\vspace*{-\baselineskip}
\end{figure}

Fig.~\ref{fig5}(a) compares photoluminescence (PL) spectra  recorded at 100~K for all samples. Each spectrum is featured by a band-edge exciton/trion (A/A$^-$) peak at $\sim$ 1.9~eV and broad luminescence (BL) peak at $\sim$ 1.7~eV\cite{pkm,tailoring,nihit1}. Deconvolution of these PL spectra by a set of six Gaussian peaks [as shown in the supplementary Fig.~S4(a)\cite{sup}] provides a good fit\cite{tailoring}. Band edge feature is consisting of two Gaussians representing free A-excitons (A) and A-trions (A$^-$), while the BL feature can be fitted with two Gaussian functions centred at 1.71~eV ($D_{hi}$) and 1.6~eV ($D_{lo}$).  $D_{hi}$ and $D_{lo}$ transitions are found to be excitonic in nature and attributed to sulfur vacancy related defects\cite{tailoring}. Note that two narrow Gaussians are used to take into account the sharp feature at $\sim$ 1.78~eV that stems from the sapphire substrate. Temperature dependent PL spectra for all samples are recorded from 20 to 300~K and deconvoluted with six Gaussians as described before. Log of the intensity associated with A-exciton [$I(A)$] is plotted as a function of inverse of temperature in fig.~\ref{fig5}(b), for various samples. In case of the reference sample R (orange), $I(A)$ increases monotonically with temperature. This tendency indicates that the excitons are activated from a bound to free sate as temperature increases. For samples deposited with 1, 5 and 7 pulses of Mn, $I(A)$ shows similar increasing tendency upto a certain temperature, beyond which it shows a reduction with rising temperature. This fall  becomes increasingly steeper  as the pulse count increases from 1 to 7. This result suggests that Mn-deposition might be introducing a non-radiative recombination channel for the A-excitons, which need to overcome a thermal barrier to get access to the non-radiative traps. Above finding further implies that the barrier height must be increasing with the amount of Mn deposited. In sample D, where Mn pulse count is 10, $I(A)$ become almost  temperature independent. In case of sample E, $I(A)$ shows similar trend as sample D [see supplementary Fig.~S4(b)\cite{sup}].  We believe that  the thermal energy barrier for the non-radiative recombination is so high for the two samples that one needs to go to higher than room temperature to see the fall of A-exciton intensity. It should be noted that in samples D and E, the density of Mn-deposition induced disorder in \mos is expected to be much more than the rest of the samples as discussed in Fig.~\ref{fig4}. At the steady state condition, density of free excitons ($n_A$)  and excitons bound to $D$ type of defect ($n_D$) should satisfy the following equations.
\begin{eqnarray}
	\frac{dn_A}{dt} &=& G - r n_A - r_n n_A -\beta (N_D - n_D) n_A + \alpha n_D = 0 \nonumber \\
	\frac{dn_D}{dt} &=& \beta (N_D - n_D) n_A -  \alpha n_D - r_D n_D  = 0 
	\label{Eq1}
\end{eqnarray}            
Where, $G$, $r$ and $r_n$ are , respectively, the rate of generation, radiative and non-radiative recombinations for the free excitons. $N_D$ is the concentration of $D$ type of defect traps and $\beta$ ($\alpha$) is the coefficient of transition of an exciton from free (defect bound) state to defect bound (free) state. $r_D$ is the recombination rate of defect bound excitons. At thermal equilibrium, $\alpha$ = $N_D$ $\beta$ $e^{-E_D/k_B T}$. $r_n$ = $r_{n0}$ $e^{-E_n/k_B T}$, where $E_D$ and $E_n$ are activation energy of excitons from defect bound to free state and the height of the thermal energy barrier for non-radiative recombination, respectively. $k_B$ is the Boltzmann constant and $r_{n0}$ is $r_n$ at $T$ = 0~K. Under the assumptions of $N_D$ $\gg$ $n_D$ and $\beta$$N_D$ $\gg$ $r_D$, the expressions for PL intensities associated with free excitons [$I(A)$= $r$ $n_A$] and the bound excitons[$I(D)$= $r_D$ $n_D$] can be obtained from eq.~\ref{Eq1} as,     
\begin{eqnarray}
	I(A) = \frac{G}{1+ \frac{r_D}{r} \exp{(\frac{E_D}{k_BT})} + \frac{r_{n0}}{r} \exp{(-\frac{E_n}{k_BT})} }
	\label{Eq2}\\
	I(D) = \frac{\frac{G r_D}{r} \exp{(\frac{E_D}{k_BT})} }{1+ \frac{r_D}{r} \exp{(\frac{E_D}{k_BT})} + \frac{r_{n0}}{r} \exp{(-\frac{E_n}{k_BT})} }
	\label{Eq3}
\end{eqnarray}

Experimentally obtained temperature variations of $I(A)$ for all samples have been fitted by eq.~\ref{Eq2}. The resultant profiles are the red solid curves shown in fig.~\ref{fig5}(b). The fit for the reference sample R returns $E_D$ = 12.4~meV and $E_n$ = 0. $E_D$ value is kept at 12.4 meV while $E_n$ is varied to fit the data for rest of the Mn-deposited samples. Obtained values for $E_n$ and $r_{n0}/r$ are plotted as functions of pulse count in fig.~\ref{fig5}(c). It can be seen that both the quantities gradually increase with the pulse count. Note that the increase of $r_{n0}/r$ could suggest the enhancement of the density of non-radiative centres in the layer. As discussed before, $E_n$ and $r_{n0}/r$ for sample D and E can not be estimated from this study.  Fig.~\ref{fig5}(d) plots logarithm of combined integrated intensities of $D_{hi}$ and $D_{lo}$ [$I(D_{hi})$ + $I(D_{lo})$], i.e., total intensity of BL feature as a function of the inverse of temperature for sample R, A, D and E. Please refer to the supplementary Fig.~S4(c)\cite{sup} for $ln[I(D_{hi})+I(D_{lo})]$ vs $1/T$ plots for samples B and C. It can be seen that like in the reference sample, in all Mn-deposited samples, intensity of this feature increases and shows a tendency of saturation with the reduction of temperature. This result implies that the Mn disorder induced non-radiative recombination channel, which modifies the variation of intensity of A-exciton with temperature in these samples, does not affect the dynamics of the sulfur deficiency bound excitonic features.  It should be noted that our attempts to simulate the data of Fig.~\ref{fig5}(d) using eq.~\ref{Eq3} with the parameters obtained from the fitting of the data using eq.~\ref{Eq2} as shown in Fig.~\ref{fig5}(b) have failed for every sample. This suggests that BL band has nothing to do with the defect D, which binds  A-excitons in our model (eq.~\ref{Eq1}). This further means that BL transition does not directly involve A-excitons. To find the binding energies associated with $D_{hi}$ and $D_{lo}$ , experimental data of [$I(D_{hi})$ + $I(D_{lo})$] vs $T$ are fitted with the equation below as described in our previous report\cite{tailoring}.
\begin{eqnarray}
	I(T)=\frac{I_0}{1+\sum \gamma_i T^{3/2} \exp(-E_{a_i}/k_B T)}
	\label{Eq4}
\end{eqnarray}
where, $E_{a_i}$ and $\gamma_i$ are the binding energy and a measure of capture cross section of $i$-th type of sulfur-vacancy related defects. Terms under  summation in the denominator take into account the contributions from the of two defect bands, $D_{lo}$ and $D_{hi}$. Temperature variation of the intensity of the defect bound BL feature is fitted with eq.~\ref{Eq4} and the resultant profiles are represented by red solid curves in fig.~\ref{fig5}(d).  Values of $E_{D_{hi}}$ ($\sim$ 100~meV) and $E_{D_{lo}}$ ($\sim$ 5~meV) are found to be almost the same for all the samples. This confirms that Mn deposition does not affect the dynamics of sulfur deficiency related BL transition in these samples. 

\begin{figure}
	\includegraphics[scale=1]{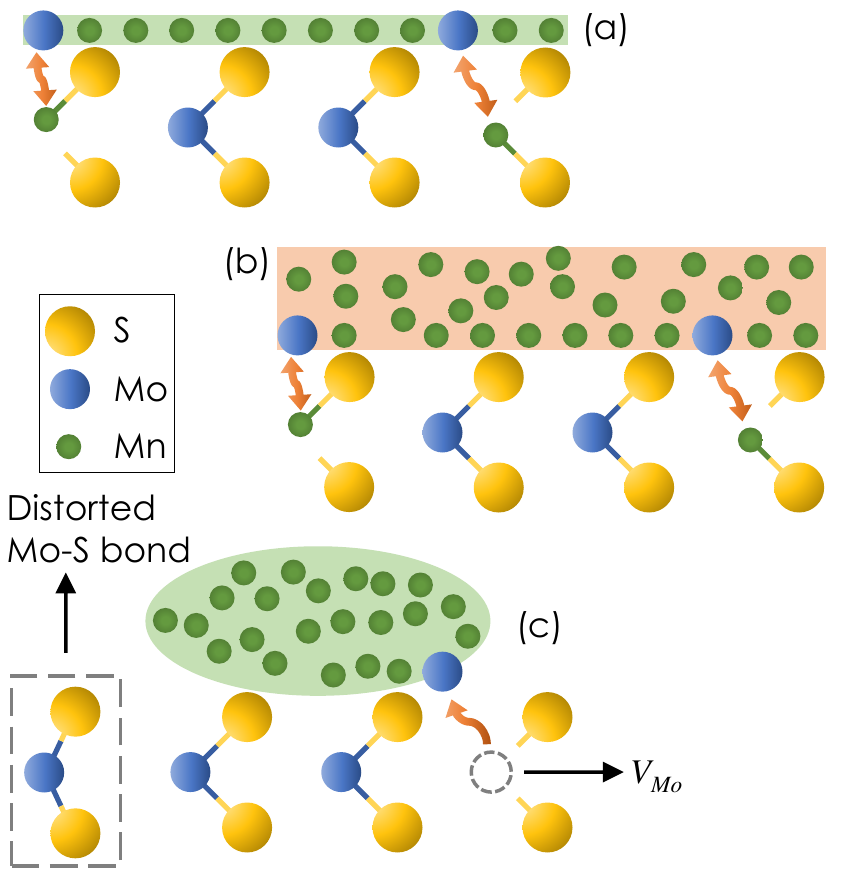}
	\caption{Schematic depiction of Mn-deposition on \mos. Island regime: (a) when Mn-deposition time is sufficiently less, flat-top Mn islands with high area coverage are present on top of \mos even after post-growth cooling of the sample to room temperature.  Droplet regime: when Mn-deposition time is sufficiently long, (b) Mn-layer with thickness more than the critical thickness is deposited on \mos. (c) However, upon post-growth-cooling of the sample to the room temperature, Mn-droplets are formed. Mo-substitution by Mn as well as the formation of Mo-vacancy (\textit{V$_{Mo}$}) and distorted Mo-S bonds are also presented in these panels.}
	\label{fig6}
%	\vspace*{-\baselineskip}
\end{figure}

From  Fig.~\ref{fig1} and \ref{fig2}  it appears that in the regime when Mn is deposited on \mos forming large flat-top islands (laser pulse count upto 5),  Mo-substitutional incorporation of Mn in 1L-\mos is more than in the regime  when Mn-droplets are formed. It is plausible that since the area coverage in the island regime is more than in the droplet regime, chance of substitution of Mo-ions by Mn-ions is expected to be more in the former case as shown schematically in Fig.~\ref{fig6}.  We believe that even in the droplet regime, Mn-layer covering a large area is deposited on \mos at the deposition temperature. A good number of Mo-substitutional incorporation of Mn is thus expected at the deposition temperature in this regime as well. However, during the post-growth-cooling stage droplets are formed in this regime. The droplet formation process give rises to either Mo-vacancies (\textit{V$_{Mo}$}) or distorted Mo-S bonds as Mn-ions are withdrawn from the substitutional sites at this stage (see Fig.~\ref{fig6}). This  can explain why the disorder in \mos increases with the pulse count, especially in the droplet regime as observed in Fig.~\ref{fig4}.  Note that reconnection of Mo with S at the Mo-substitutional site upon withdrawal of Mn during droplet formation stage seems to be quite likely in the present case as the relative strength of the knocked out Mo 3$d_{5/2}$ signal with respect to Mo-S signal in Fig.~\ref{fig2}(k) dies off for more than 5 pulse counts of Mn deposition. Results of Fig.~\ref{fig5} show that the non-radiative  recombination centres for the excitons in \mos films increase with  Mn-deposition time (pulse count).  Interestingly, the thermal energy barrier for such recombinations $E_n$ has also been found to increase with pulse count.  It is plausible that Mo-related defects act as the non-radiative centres, each one of which offers a particular value of  $E_n$ around it. However, a cluster of such defects may offer an entirely different barrier height.  We believe that the barrier height $E_n$ increases with the cluster size in these samples. Since the probability of formation of bigger clusters  enhances with the concentration of Mo-defects, $E_n$ rises with the pulse count.

%\section{Conclusion}
The study  shows that Mo-substitutional incorporation of Mn in 1L-\mos is possible when it is deposited below a certain critical amount on the films using PLD technique. In this deposition regime, thin Mn-coating covering large areas can takes place on \mos layers. Large area coverage is believed to enhance the chance of substitutional incorporation of Mn. However, dewetting occurs when Mn-deposition is more than the critical amount, which leads to the emergence of Mn-droplets on \mos layers. Due to the reduction of the coverage area, substitutional incorporation of Mn is suppressed in this regime. On the other hand,  disorder in the lattice is observed to be enhanced with the deposition time.  Mn-deposition is also found to influence strongly the excitonic recombination dynamics of 1L-\mos by introducing non-radiative recombination centres for the excitons. The density of these centres as well as the barrier height for such recombinations  to happen increase rapidly with the amount of Mn-deposition, especially in the droplet-formation regime.  These non-radiative transitions can be attributed to  either Mo-vacancies or distorted Mo-S bonds. A large number of such defects are believed to be produced during the formation of the Mn-droplets as a result of the withdrawal of Mn ions from the Mo-substitutional sites.
\\ 
\vspace*{-\baselineskip}

%\noindent\textbf{Supplementary:}
%See supplementary information for additional information regarding AFM, XPS, Raman and PL %studies on Mn deposited 1L-{MoS$_2$} samples.

\noindent\textbf{Acknowledgment:}
We acknowledge the financial support by Department of Science and Technology (DST) of Government of India under the grant number: CRG/2018/001343. We would like to thank Industrial Research and Consultancy Centre (IRCC), Sophisticated Analytical Instrument Facility (SAIF), and Centre of Excellence in Nanoelectronics (CEN) of IIT Bombay, for providing various experimental facilities. 

\bibliographystyle{unsrt}
\bibliography{BIBManuscript}

\begin{thebibliography}{10}

\bibitem{mak1}
Kin~Fai Mak, Changgu Lee, James Hone, Jie Shan, and Tony~F. Heinz.
\newblock Atomically thin {MoS$_2$} a new direct-gap semiconductor.
\newblock {\em Phys. Rev. Lett.}, 105:136805, 2010.

\bibitem{mak2}
Kin~Fai Mak, Keliang He, Jie Shan, and Tony~F Heinz.
\newblock Control of valley polarization in monolayer {MoS2} by optical
  helicity.
\newblock {\em Nature Nanotechnology}, 7(8):494--498, August 2012.

\bibitem{VP}
Ting Cao, Gang Wang, Wenpeng Han, Huiqi Ye, Chuanrui Zhu, Junren Shi, Qian Niu,
  Pingheng Tan, Enge Wang, Baoli Liu, and Ji~Feng.
\newblock Valley-selective circular dichroism of monolayer molybdenum
  disulphide.
\newblock {\em Nature Communications}, 3(1):887, June 2012.

\bibitem{VP1}
Hualing Zeng, Junfeng Dai, Wang Yao, Di~Xiao, and Xiaodong Cui.
\newblock Valley polarization in {MoS2} monolayers by optical pumping.
\newblock {\em Nature Nanotechnology}, 7(8):490--493, August 2012.

\bibitem{bs}
Andrea Splendiani, Liang Sun, Yuanbo Zhang, Tianshu Li, Jonghwan Kim, Chi-Yung
  Chim, Giulia Galli, and Feng Wang.
\newblock Emerging photoluminescence in monolayer {MoS$_2$}.
\newblock {\em Nano Lett.}, 10:1271, 2010.

\bibitem{3D_flex}
Ruiping Zhou and Joerg Appenzeller.
\newblock Three-dimensional integration of multi-channel mos<inf>2</inf>
  devices for high drive current fets.
\newblock In {\em 2018 76th Device Research Conference (DRC)}, pages 1--2,
  2018.

\bibitem{magnet1}
Y.~C. Cheng, Z.~Y. Zhu, W.~B. Mi, Z.~B. Guo, and U.~Schwingenschl\"ogl.
\newblock Prediction of two-dimensional diluted magnetic semiconductors: Doped
  monolayer mos${}_{2}$ systems.
\newblock {\em Phys. Rev. B}, 87:100401, Mar 2013.

\bibitem{magnet2}
Ashwin Ramasubramaniam and Doron Naveh.
\newblock Mn-doped monolayer mos${}_{2}$: An atomically thin dilute magnetic
  semiconductor.
\newblock {\em Phys. Rev. B}, 87:195201, May 2013.

\bibitem{magnet3}
Qu~Yue, Shengli Chang, Qin Shiqiao, and Jingbo Li.
\newblock Functionalization of monolayer mos2 by substitutional doping: A
  first-principles study.
\newblock {\em Physics Letters A}, 377:1362–--1367, 08 2013.

\bibitem{VP2}
Di~Xiao, Gui-Bin Liu, Wanxiang Feng, Xiaodong Xu, and Wang Yao.
\newblock Coupled spin and valley physics in monolayers of ${\mathrm{mos}}_{2}$
  and other group-vi dichalcogenides.
\newblock {\em Phys. Rev. Lett.}, 108:196802, May 2012.

\bibitem{XPS1}
Jeffrey~R. Lince, Thomas~B. Stewart, Paul~D. Fleischauer, Jory~A. Yarmoff, and
  Amina Taleb‐Ibrahimi.
\newblock The chemical interaction of mn with the {MoS$_2$}(0001) surface
  studied by high‐resolution photoelectron spectroscopy.
\newblock {\em Journal of Vacuum Science \& Technology A}, 7:2469--2474, 1989.

\bibitem{ch1}
Shinichiro Mouri, Yuhei Miyauchi, and Kazunari Matsuda.
\newblock Tunable photoluminescence of monolayer mos2 via chemical doping.
\newblock {\em Nano Letters}, 13(12):5944--5948, 2013.

\bibitem{ch2}
Goki Eda, Hisato Yamaguchi, Damien Voiry, Takeshi Fujita, Mingwei Chen, and
  Manish Chhowalla.
\newblock Photoluminescence from chemically exfoliated mos2.
\newblock {\em Nano Letters}, 11(12):5111--5116, 2011.

\bibitem{ch3}
Matin Amani, Der-Hsien Lien, Daisuke Kiriya, Jun Xiao, Angelica Azcatl, Jiyoung
  Noh, Surabhi~R. Madhvapathy, Rafik Addou, Santosh KC, Madan Dubey, Kyeongjae
  Cho, Robert~M. Wallace, Si-Chen Lee, Jr-Hau He, Joel~W. Ager, Xiang Zhang,
  Eli Yablonovitch, and Ali Javey.
\newblock Near-unity photoluminescence quantum yield in mos<sub>2</sub>.
\newblock {\em Science}, 350(6264):1065--1068, 2015.

\bibitem{therm}
S.~Deb, P.~Bhattacharyya, P.~Chakrabarti, H.~Chakraborti, K.~Das Gupta,
  A.~Sukla, and S.~Dhar.
\newblock Effect of oxygen adsorption on electrical and thermoelectrical
  properties of monolayer {MoS$_2$}.
\newblock {\em Phys. Rev. Applied}, 14:034030, 2020.

\bibitem{ad2}
Rahul Rao, Victor Carozo, Yuanxi Wang, Ahmad~E Islam, Nestor Perea-Lopez,
  Kazunori Fujisawa, Vincent~H Crespi, Mauricio Terrones, and Benji Maruyama.
\newblock Dynamics of cleaning, passivating and doping monolayer {MoS}$_2$ by
  controlled laser irradiation.
\newblock {\em 2D Materials}, 6:045031, 2019.

\bibitem{ad3}
Pranjal~Kumar Gogoi, Zhenliang Hu, Qixing Wang, Alexandra Carvalho, Daniel
  Schmidt, Xinmao Yin, Yung-Huang Chang, Lain-Jong Li, Chorng~Haur Sow,
  A.~H.~Castro Neto, Mark B.~H. Breese, Andrivo Rusydi, and Andrew T.~S. Wee.
\newblock Oxygen passivation mediated tunability of trion and excitons in
  {MoS$_2$}.
\newblock {\em Phys. Rev. Lett.}, 119:077402, 2017.

\bibitem{ad4}
Sefaattin Tongay, Jian Zhou, Can Ataca, Jonathan Liu, Jeong~Seuk Kang, Tyler~S.
  Matthews, Long You, Jingbo Li, Jeffrey~C. Grossman, and Junqiao Wu.
\newblock Broad-range modulation of light emission in two-dimensional
  semiconductors by molecular physisorption gating.
\newblock {\em Nano Letters}, 13:2831, 2013.

\bibitem{evap1}
Hui Fang, Mahmut Tosun, Gyungseon Seol, Ting~Chia Chang, Kuniharu Takei, Jing
  Guo, and Ali Javey.
\newblock Degenerate n-doping of few-layer transition metal dichalcogenides by
  potassium.
\newblock {\em Nano Letters}, 13(5):1991--1995, 2013.

\bibitem{raman2}
Nihit Saigal, Isabelle Wielert, Davor Čapeta, Nataša Vujičić, Boris~V.
  Senkovskiy, Martin Hell, Marko Kralj, and Alexander Grüneis.
\newblock Effect of lithium doping on the optical properties of monolayer
  {MoS$_2$}.
\newblock {\em Applied Physics Letters}, 112:121902, 2018.

\bibitem{ion1}
Sefaattin Tongay, Joonki Suh, Can Ataca, Wen Fan, Alexander Luce, Jeong~Seuk
  Kang, Jonathan Liu, Changhyun Ko, Rajamani Raghunathanan, Jian Zhou, Frank
  Ogletree, Jingbo Li, Jeffrey~C. Grossman, and Junqiao Wu.
\newblock Defects activated photoluminescence in two-dimensional
  semiconductors: interplay between bound, charged, and free excitons.
\newblock {\em Sci. Rep.}, 3:2657, 2013.

\bibitem{raman1}
Sandro Mignuzzi, Andrew~J. Pollard, Nicola Bonini, Barry Brennan, Ian~S.
  Gilmore, Marcos~A. Pimenta, David Richards, and Debdulal Roy.
\newblock Effect of disorder on raman scattering of single-layer
  $\mathrm{Mo}{\mathrm{s}}_{2}$.
\newblock {\em Phys. Rev. B}, 91:195411, May 2015.

\bibitem{XPS2}
Kehao Zhang, Simin Feng, Junjie Wang, Angelica Azcatl, Ning Lu, Rafik Addou,
  Nan Wang, Chanjing Zhou, Jordan Lerach, Vincent Bojan, Moon~J. Kim, Long-Qing
  Chen, Robert~M. Wallace, Mauricio Terrones, Jun Zhu, and Joshua~A. Robinson.
\newblock Manganese doping of monolayer mos2: The substrate is critical.
\newblock {\em Nano Letters}, 15(10):6586--6591, 2015.

\bibitem{pkm}
P.~K. Mohapatra, S.~Deb, B.~P. Singh, P.~Vasa, and S.~Dhar.
\newblock Strictly monolayer large continuous {MoS$_2$} films on diverse
  substrates and their luminescence properties.
\newblock {\em Appl. Phys. Lett.}, 108:042101, 2016.

\bibitem{tailoring}
S.~Deb, P.~Chakrabarti, P.~K. Mohapatra, B.~K. Barick, and S.~Dhar.
\newblock Tailoring of defect luminescence in cvd grown monolayer {MoS$_2$}
  film.
\newblock {\em Applied Surface Science}, 445:542--547, 2018.

\bibitem{sup}
See supplementary material for additional information regarding AFM, XPS, Raman
  and PL studies on Mn deposited 1L-MoS$_2$ samples.

\bibitem{XPS}
H~W Wang, P~Skeldon, and G~E Thompson.
\newblock {XPS} studies of {MoS$_2$} formation from ammonium tetrathiomolybdate
  solutions.
\newblock {\em Surf. Coat. Technol.}, 91:200, 1997.

\bibitem{XPS3}
C.~J. Jenks, S.~L. Chang, J.~W. Anderegg, P.~A. Thiel, and D.~W. Lynch.
\newblock Photoelectron spectra of an
  ${\mathrm{al}}_{70}$${\mathrm{pd}}_{21}$${\mathrm{mn}}_{9}$ quasicrystal and
  the cubic alloy ${\mathrm{al}}_{60}$${\mathrm{pd}}_{25}$${\mathrm{mn}}_{15}$.
\newblock {\em Phys. Rev. B}, 54:6301--6306, 1996.

\bibitem{XPS5}
Y~Umezawa and C~N Reilley.
\newblock Effect of argon ion bombardment on metal complexes and oxides studied
  by x-ray photoelectron spectroscopy.
\newblock {\em Anal. Chem.}, 50:1290, 1978.

\bibitem{XPS4}
Brian~R. Strohmeier and David~M. Hercules.
\newblock Surface spectroscopic characterization of manganese/aluminum oxide
  catalysts.
\newblock {\em J. Phys. Chem.}, 88(21):4922--4929, 1984.

\bibitem{raman4}
Marcel Placidi, Mirjana Dimitrievska, Victor Izquierdo-Roca, Xavier
  Fontan{\'{e}}, Andres Castellanos-Gomez, Amador P{\'{e}}rez-Tom{\'{a}}s,
  Narcis Mestres, Moises Espindola-Rodriguez, Simon L{\'{o}}pez-Marino, Markus
  Neuschitzer, Veronica Bermudez, Anatoliy Yaremko, and Alejandro
  P{\'{e}}rez-Rodr{\'{\i}}guez.
\newblock Multiwavelength excitation raman scattering analysis of bulk and
  two-dimensional {MoS$_2$}:vibrational properties of atomically thin {MoS$_2$}
  layers.
\newblock {\em 2D Materials}, 2(3):035006, 2015.

\bibitem{raman5}
Bruno~R Carvalho, Yuanxi Wang, Sandro Mignuzzi, Debdulal Roy, Mauricio
  Terrones, Cristiano Fantini, Vincent~H Crespi, Leandro~M Malard, and Marcos~A
  Pimenta.
\newblock Intervalley scattering by acoustic phonons in two-dimensional
  {MoS$_2$} revealed by double-resonance raman spectroscopy.
\newblock {\em Nature Communications}, 8(1):14670, 2017.

\bibitem{raman3}
Biswanath Chakraborty, Achintya Bera, D.~V.~S. Muthu, Somnath Bhowmick, U.~V.
  Waghmare, and A.~K Sood.
\newblock Symmetry-dependent phonon renormalization in monolayer {MoS$_2$}
  transistor.
\newblock {\em Phys. Rev. B}, 85:161403, 2012.

\bibitem{nihit1}
Nihit Saigal and Sandip Ghosh.
\newblock Evidence for two distinct defect related luminescence features in
  monolayer {MoS$_2$}.
\newblock {\em Appl. Phys. Lett.}, 109:122105, 2016.

\end{thebibliography}

%\end{document}
%
% ****** End of file apstemplate.tex ******

%%%%%%%%%%%%%%%%%%%%%%%%%%%%%%%%%%%%%%%%%%%%%%%%%%%%%%%%%%%%
%\newcommand{\taum}{\tau_\text{\tiny M}}
%\documentclass[journal]{paper}
%%%%%%%%%%%%%%%%%%%%%%%%%%%%%%%%%%%%%%%%%%%%%%%%%%%%%%%%%%%
%\usepackage{xspace} 
%\usepackage{cite} %1-3,5-7 instead 1,2,3,5,6,7
%\usepackage{makecell} %\thead
%\usepackage{siunitx, mhchem}
%\usepackage{color,soul}
%\usepackage{graphicx}
%\usepackage{bm}        % for math
%\usepackage{amssymb}   % for math
%\usepackage{amsfonts}
%\usepackage{amsmath,mathtools} % mathtools to use \begin{multlined}
%\usepackage{breqn} %Break Equation automatically
%\usepackage{multirow}
%\usepackage{array}
%\usepackage{booktabs}
%\usepackage[export]{adjustbox}
%\usepackage{textcomp}%for \textdegree
%\usepackage[numbers]{natbib}
%\usepackage{caption}
%\renewcommand{\thefigure}{{\small S}\arabic{figure}}
%\usepackage{chngcntr}
%\renewcommand{\thetable}{{\small S}\arabic{table}}
%%%%%%%%%%%%%%%%%%%%%%%%%%%%%%%%%%%%%%%%%%%%%%%%%%%%%%%%%%%%
%\newcommand{\mos}{MoS$_2$\xspace}
%\newcommand{\ws}{WS$_2$ }
%\newcommand{\degree}{$^{\small{\text{o}}}$}
%%%%%%%%%%%%%%%%%%%%%%%%%%%%%%%%%%%%%%%%%%%%%%%%%%%%%%%%%%%%

%\begin{document}

% Use the \preprint command to place your local institutional report
% number in the upper righthand corner of the title page in preprint mode.
% Multiple \preprint commands are allowed.
% Use the 'preprintnumbers' class option to override journal defaults
% to display numbers if necessary
%\preprint{}

%Title of paper

\noindent
{\huge Supplementary Information}
\\

\noindent
\textbf{{\Large{Effect of manganese incorporation on the excitonic recombination dynamics in monolayer MoS$_2$}}}
%\title{{\Large Supplementary Information}\\
%Effect of manganese incorporation on the excitonic recombination dynamics in %monolayer \mos}
\\
{\large Poulab Chakrabarti, Santosh Kumar Yadav, Swarup Deb, Subhabrata Dhar}\\
Department of Physics, Indian Institute of Technology Bombay, Powai, Mumbai-400076, India
%\\

%\author{Poulab\ Chakrabarti*, Santosh\ Kumar\ Yadav, Swarup\ Deb, Subhabrata\ Dhar\\
%{\normalsize Department of Physics, Indian Institute of Technolagy Bombay, %Mumbai-400076,}\\{\normalsize India}\\
%{\normalsize \texttt{$^*$E-mail:~poulab007@gmail.com}}}

\date{\today}
\maketitle

\clearpage

\begin{figure}
	\includegraphics[scale=1.7]{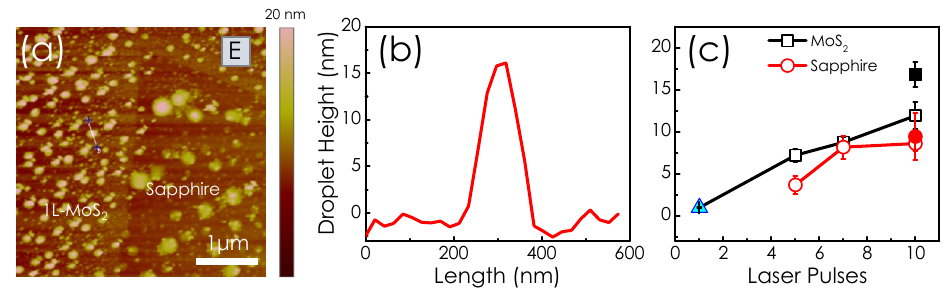}
	\caption{Panel (a) is showing AFM topography image of sample E where Mn is deposited for 10 counts of laser pulses at 8.0$\times$10$^{-5}$ mbar chamber pressure. Mn droplets on both 1L-\mos and sapphire surfaces are evident. Droplet density is found to be maximum for this sample. (b) Shows the height profile of one of the droplet formed on \mos. (c) Plots the average height of the Mn droplets forming on 1L-\mos (black) and sapphire (red) portions of all Mn-deposited samples. Filled symbols at 10 pulses represent sample E. Blue triangle represents the average height of the isolated Mn islands on top of \mos for sample A, where Mn is deposited for only 1 pulse (see fig.~1~(a) of the main text).}
\end{figure}
\clearpage
\begin{figure}[h!]
	\includegraphics[scale=1.4]{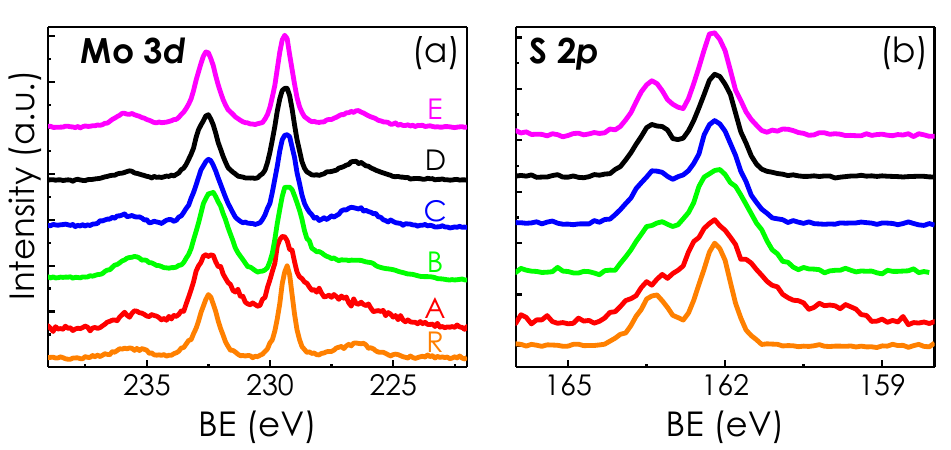}
	\caption{(a) XPS spectra associated with Mo $3d$ and S $2s$ levels for all the samples. Spectra are normalized with respect to $\sim$ 229.3~eV Mo $3d_{5/2}$ peak. Evidently, all the features (including the S $2s$ peak) are broader for sample A (1 pulse) and B (5 pulses) as compared to those of the reference sample R. Broadening is a result of substitution of Mo by Mn in \mos lattice as discussed in the main text. Note that the broadening of these peaks is again decreased in sample C(7 pulses), D(10 pulses) and E(10 pulses). (b) Normalized (at $\sim$ 162.2~eV S $2p_{3/2}$ peak) XPS spectra associated with S $2p$ core levels for all samples. Like Mo $3d$ peaks, these features also show the enhancement of broadening for sample A and B as compared to other samples.}
	\label{Fig:2}
\end{figure}

\begin{figure}
	\includegraphics[scale=1.7]{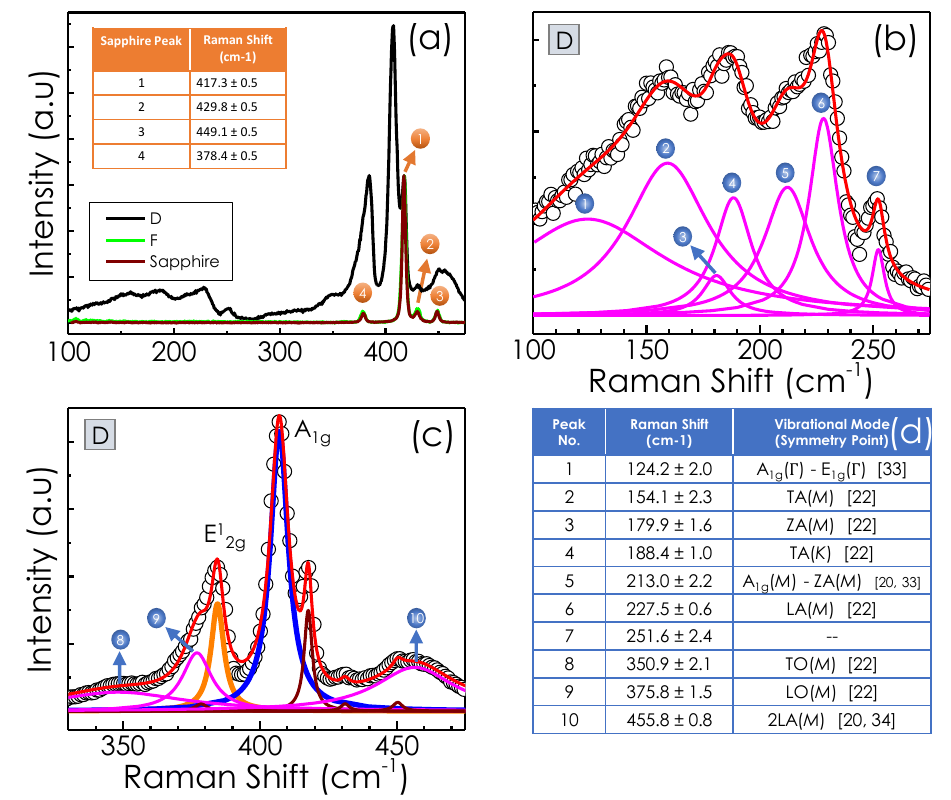}
	\caption{(a)Raman spectra for sample D (black), F (Green) and a bare sapphire substrate (brown). Inset shows a table where the positions of the four sapphire peaks are provided. Spectral resolution of the Raman spectrometer, which is $\sim$ 0.5 ~cm$^{-1}$, is taken as the error bar in the peak-position. These three spectra are normalized at $\sim$ 417.3~cm$^{-1}$, which is the position of most intense sapphire peak. Evidently there is no difference between the Raman spectra of the bare sapphire substrate and sapphire coated with Mn  (5 pulses) in sample F, demonstrating that Mn deposition does not result in any additional Raman feature. Deconvolution of the (b) low and (c) high frequency part of the Raman spectra for sample D (10 pulses). Scatter plots are the experimental data, red curve represent the sum of all deconvoluted peaks, orange and blue lines represent E$^1_{2g}$  and A$_{1g}$ modes of \mos,  brown peaks are stemming  from sapphire substrate and magenta peaks are defect related peaks of \mos, which are marked by blue numbers. These peaks are identified following previous works in the table shown in panel (d)\cite{raman1, raman2, raman4, raman5}.}
\end{figure}

\begin{figure}
	\includegraphics[scale=1.4]{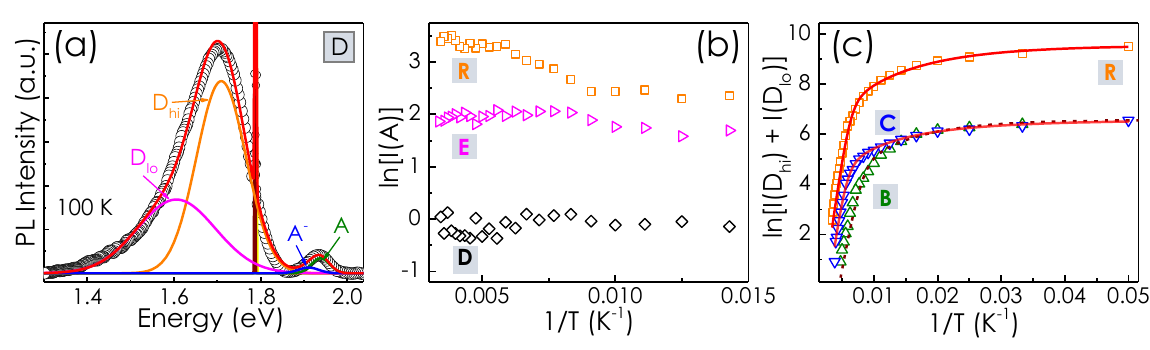}
	\caption{(a) Deconvolution of PL spectrum for sample D at 100~K temperature, by a set of six Gaussians\cite{tailoring}. Scatter plot is the experimental data while the solid red line represents the fitting. Two Gaussians at 1.93~eV (green line) and 1.91~eV (blue line) represent band-edge free exciton (A) and A-trion (A$^-$). Two sapphire peaks at $\sim$ 1.78~eV are taken care of by two sharp peaks (yollow and brown lines). The broad luminescence (BL) peak at $\sim$1.7~eV is composed of two Gaussian features centered at 1.71~eV ($D_{hi}$) and 1.6~eV ($D_{lo}$). In one of our previous works\cite{tailoring}, the two transitions ($D_{hi}$ and $D_{lo}$) are shown to be excitonic in nature and are resulting from sulfur vacancy related defects. (b) Logarithm of the intensity contribution of the A-exciton [$I(A)$] versus inverse of temperature ($1/T$) for samples R (orange), D  (black) and E (magenta). While $ln[I(A)]$ clearly increases with temperature for sample R, it does not vary much throughout the temperature range for sample D and E. (c) Logarithm of total integrated PL intensity of BL feature [$I(D_{hi})$ + $I(D_{lo})$] versus $1/T$ for samples R (orange), B (green) and C (blue). The solid lines represent the fit using eq.~4 of the manuscript.}
\end{figure}

%\clearpage
%\bibliographystyle{unsrt}
%\bibliography{bibsup}
%\end{document}

\end{document}